\documentclass[prb,aps,twocolumn,groupedaddress,floats,final,nopacks]{revtex4-2}
\usepackage{graphicx}
\usepackage{subfigure}
\usepackage{amssymb}
\usepackage{latexsym}
\usepackage{epsfig}
\usepackage{psfrag}
\usepackage{dcolumn}
\usepackage{amsmath,amsfonts,amssymb,bm,ulem}
\usepackage{color}
\usepackage{float}

\newcommand{\be}{\begin{equation}}
\newcommand{\ee}{\end{equation}}
\newcommand{\bea}{\begin{eqnarray}}
\newcommand{\eea}{\end{eqnarray}}

\begin{document}
\title{Dielectric losses in metals}

\author{I.S. Tupitsyn}
\affiliation{Department of Physics, University of Massachusetts, Amherst, MA 01003, USA}
\author{N.V. Prokof'ev}
\affiliation{Department of Physics, University of Massachusetts, Amherst, MA 01003, USA}

%\date{\today}

\begin{abstract}
Bethe-Salpeter equation (BSE) in the self-consistent Hartree-Fock (HF) basis is often used for describing complex many-body effects in material science applications. Its exact solution on the real-frequency axis at finite temperature for polarization using the diagrammatic Monte Carlo method [Phys. Rev. B \textbf{109}, 045152, (2024)] captures effects of multiple Coulomb scattering of a single particle-hole excitation, but does not account for multiple pair excitations important for studying dielectric loses in metals at frequencies comparable to the plasmon mode. In this paper we report technical developments which allow one to efficiently compute the dielectric response in a wide frequency range from zero to a few Fermi energies without systematic bias at finite $T$. By applying it to the homogeneous electron gas we demonstrate how at small momenta the gap in the spectral density between the electron-hole and plasmon excitations, existing within the HF-BSE approach, is filled with two particle-hole excitations and is completely washed out already at temperature $T \sim \varepsilon_F/10$.
\end{abstract}

\maketitle
%%%%%%%%%%%%%%%%%%%%%%%%%%%%%%%%%%%%%%%%%%%%%%%%%%%%%%%%%%%%%%%%%%%%%%%%%%%%%%%%%%

\noindent \textit{Introduction.} The so-called energy loss function (ELF), $-{\rm Im} \; \epsilon^{-1}$, where $\epsilon$ is the dielectric function, is a key property responsible for dissipative processes in metals. It contains information not only about single particle-hole and plasmon excitations, but also about multiple excitations which dominate the answer in regions where otherwise one would find a spectral gap. The most relevant fundamental model featuring the physics of electron liquid in metals is the homogeneous electron gas (HEG) defined by the Hamiltonian
\begin{equation}
H=\sum_i \frac{k_i^2}{2m} + \sum_{i<j} \frac{e^2}{|\mathbf{r}_i -\mathbf{r}_j|} - \mu N,
\label{jellium}
\end{equation}
or jellium \cite{Ceperley80,Utsumi80,Perdew92}, with studies of its dielectric losses due to interactions between the charges going back to the work of Lindhard and Winther \cite{Lindhard64}. Here $m$ and $e$ are the electron mass and charge and $\mu$ is the chemical potential. In what follows we measure momenta in units of the Fermi momentum $k_F$ and energies in units of the Fermi energy $\varepsilon_F = k_F^2$; the gas density is thus kept fixed at $\rho = k_F^3/3\pi^2$, which is achieved by fine-tuning the chemical potential. The Coulomb parameter $r_S=(4\pi a_B^3 \rho / 3)^{-1/3}$ characterizing the strength of many-body correlations (typical values of $r_s$ in metals are in the $r_S=2 \div 5$ range) measures the inter-particle distance in terms of the Bohr radius, $a_B=1/me^2$.

The dielectric function, $\epsilon(Q,\Omega,T)$, dependence on momentum, $Q$, frequency, $\Omega$, and temperature, $T$, is complicated and its accurate calculation is a very challenging theoretical task. It can be expressed through
the system's polarization as $\epsilon = 1 - V \Pi$, where $V(Q) = 4 \pi e^2 / Q^2$ is the Coulomb potential,
and computed using field theoretical techniques for polarization $\Pi$. In the simplest possible random-phase approximation (RPA) \cite{Lindhard54} the polarization is estimated from the convolution of two bare Green's functions (the lowest order one-loop Feynman diagram); in the Hedin's skeleton approach this diagram is a part of the self-consistent GW-approximation \cite{hedin65,hedin69}. However, RPA is guaranteed to provide accurate results only in the limit of small Coulomb parameter $r_S$, while the fully self-consistent GW-approximation violates the exact hydrodynamic condition, $\Pi (Q\to 0, \Omega \ne 0) \propto Q^2$ \cite{Pines,Holm98,JelGW} and eliminates the
plasmon mode at small momenta \cite{TTKP2021}. Applications of the time-dependent density functional theory (TDDFT) to this problem are also limited to approximations involving static exchange-correlation kernels missing effects of excited states, or phenomenological kernels missing proper frequency dependence \cite{vignale} (see also discussion in Ref.~\cite{LCHPT2022}). This motivates the development of computational tools that allow one to compute contributions to $\Pi$ from higher-order Feynman diagrams, e.g. by accounting for multiple Coulomb scattering between the excited particle and hole at the level of Bethe-Salpeter equation (BSE) in the self-consistent Hartree-Fock (HF) basis and adding multi-loop contributions.

The ill-conditioned procedure (unless explicitly biased by imposed constraints) of obtaining real-frequency results from imaginary frequency/time data was eliminated by formulation and development of the algorithmic Matsubara intergation (AMI) technique in Refs.~\cite{LeBlanc2019,LeBlanc2020a,LeBlanc2020b} and its generalization in Ref.~\cite{TTKP2021}. It is based on the observation that all Matsubara sums and the Wick's rotation of external frequency, $i\Omega \to \Omega +i\eta$, can be done analytically for an arbitrary Feynman diagram. With finite regularization parameter $\eta >0$ this technique was used to compute the dielectric function and exchange-correlation kernel of the HEG in Ref.~\cite{LCHPT2022}. Finite $\eta$ is required to avoid divergent statistical measures and variance when performing stochastic sampling of functions featuring multiple poles without implementing additional analytic transformations of the integrand in the vicinity of poles. On the one hand, large $\eta $ introduces systematic bias and may radically change the result in the spectral gap region, especially at low temperature. On the other hand, small values of $\eta $ make calculations prohibitively expensive numerically, especially for diagrams featuring poles of second (and higher) order.

Mathematically, the problem of pole regularization can be formulated as follows: given an integrand featuring a pole of order $n$
\begin{equation}
I_n = \int^{\infty}_{-\infty} dx  R_n(x); \;\; R_n(x)=\frac{h(x)}{(\Omega-g(x)+i\eta)^n},
\label{In}
\end{equation}
find the transformation $R_n(x) \to \bar{R}_n(x)$ such that (i) the integral does not change, and (ii) $ \bar{R}_n(x)$ function is not singular at the pole $x=x_0$, where $g(x_0)=\Omega$. By replacing $R_n$ with $\bar{R}_n$ one explicitly takes the limit of $\eta \to 0$ and obtains formulation suitable for efficient Monte Carlo evaluation of the integral.

\begin{figure}[t]
\centerline{\includegraphics[width=0.95\columnwidth]{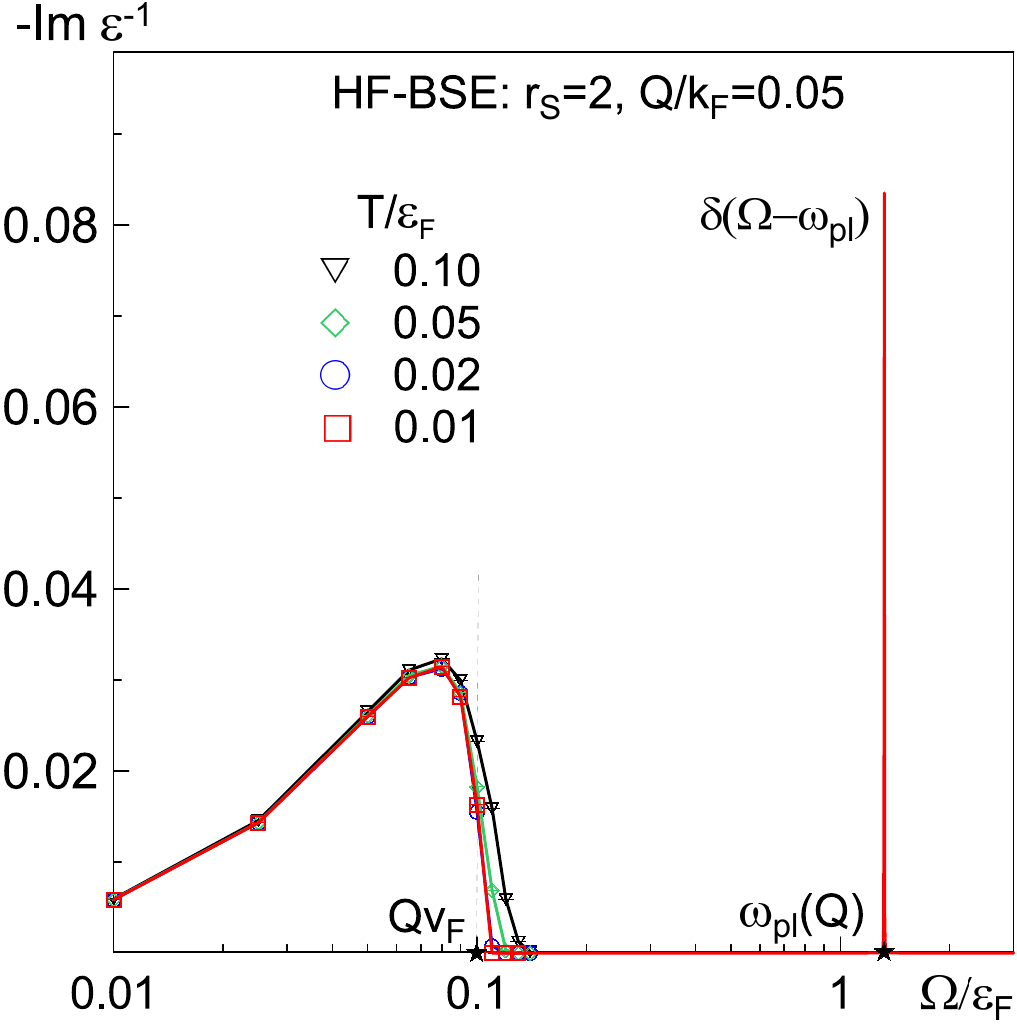}}
\caption{Energy loss function at $r_S=2$ and $Q/k_F=0.05$ within the HF-BSE scheme of Ref.~\cite{BSE2024}.
Curves with symbols correspond to different temperatures.
The vertical solid red line at $\Omega/\varepsilon_F \approx 1.33$ is the $\delta$-functional plasmon peak.
Error bars are within the symbol sizes.}
\label{Fig1}
\end{figure}

For first-order poles the corresponding procedure was explained in Ref.~\cite{TTKP2021} and is nothing but the well-known decomposition into $\delta$-functional and principal value parts:
\begin{equation}
\rm{Im} \bar{R}_1 = - \pi \frac{ h(x_0)}{|g'(x_0)|} N(x,x_0) \,
\label{R1a}
\end{equation}
\begin{equation}
\rm{Re} \bar{R}_1(x) =\left\{
\begin{array}{ll}
         [R_1(x)  + R_1(x_r) ]/2   & \mbox{if}\; |x-x_0|<a \\
             R_1(x)                & \mbox{otherwise}
\end{array} \right.
\label{R1b}
\end{equation}
where $x_r=2x_0-x$ and $N(x,x_0)$ is an arbitrary function  normalized to unity on the $(x_0-a,x_0+a)$ and zero
otherwise. Formally the value of $a>0$ is arbitrary, but by keeping it relatively small one can use $\bar{R}_1=R_1$ away from poles. [In simulations performed in this work, poles were regularized using $x = \mathbf{k} \cdot {\mathbf Q}/Q$ as an integration variable and $a=k_F/4$.] Implementation of this protocol for simple poles already allows one to achieve numerically exact solution for polarization within the HF-BSE scheme \cite{BSE2024,LD2024}. In Fig.~\ref{Fig1} we show ELF results for the HEG at $r_S=2$ and different temperatures based on this approximation. The low frequency peak is due to single electron-hole (e-h) pair excitations while the $\delta$-functional peak at $\omega_{pl}$ corresponds to the plasmon mode. Finite temperature effects appear modest even at $T/\varepsilon_F =0.1$ and limited to broadening of the e-h continuum threshold at $\omega \sim Qv_F$, where $v_F$ is the Fermi velocity. A nearly perfect (up to exponentially small thermal component) spectral gap between the e-h continuum and the plasmon peak, undamped plasmon mode, and near-zero spectral density at $\Omega > \omega_{pl}$ remain unchanged as they are characteristic for both the RPA and HF-BSE approximations; the most important improvement achieved by the latter is substantial renormallization of the Landau damping coefficient \cite{LD2024}.

The goal of this work is to present solutions to problem (\ref{In}) for $n>1$, which
is the major technical obstacle preventing efficient bias-free Monte Carlo simulations of diagrams
featuring high-order poles, and use them to compute contributions to the ELF in regions where HF-BSE
approximation has spectral gaps. The accuracy of the proposed diagrammatic setup is verified by
comparison with the finite-$\eta$ scheme of Ref.~\cite{LCHPT2022} at the highest simulated temperature
when the effect of $\eta$ can be made negligible.

%%%%%%%%%%%%%%%%%%%%%%%%%%%%%%%%%%%%%%%%%%%%%%%%%%%%%%%%%%%%%%%%%%%%%
\smallskip

\noindent \textit{Regularization of poles.}
Given Eqs.~(\ref{R1a}) and (\ref{R1b}) let us proceed with regularization of higher-order poles. Imaginary part is the easiest to deal with because it always comes from the singular behavior of the integrand at the pole location $x=x_0$. Since $I_2 = - dI_1/d\Omega$, one can directly differentiate Eq.~(\ref{R1a}) using $d\Omega=g'(x_0)dx_0$ to get
\begin{equation}
\rm{Im} \bar{R}_2 = \pi \frac{g'_0 h'_0-g''_0 h_0}{|g'_0|^3} \, N.
\label{R2a}
\end{equation}
To simplify notations, we used subscript ``0'' for values of all functions at $x=x_0$ and omitted arguments of the arbitrary normalized function $N \equiv N(x,x_0)$. This trick can be repeated to obtain explicit expressions for imaginary parts of $R_n$ for any $n$. For example, using $I_3 = - (1/2) dI_2/d\Omega$ one obtains
\begin{equation}
\rm{Im} \bar{R}_3 = \frac{\pi}{2} \frac{ [g'_0g'''_0-3(g''_0)^2] h_0 + 3 g'_0 g''_0 h'_0 - (g'_0)^2 h''_0 }
{ |g_0'|^5}\, N.
\label{R3a}
\end{equation}

Real part regularization is based on observation that for any $\eta > 0$
\begin{equation}
\int_{-\infty}^{\infty} {\rm Re}\frac{1}{(i\eta -z )^{2}} dz =
 \frac{1}{\eta} \int_{-\infty}^{\infty} dz  \frac{z^2-1}{ (z^2 + 1)^2 } = 0.
\label{Reg1}
\end{equation}
If the integration domain is finite, then
\begin{equation}
0 = \int_{-a}^{a} dz  {\rm Re}\frac{1}{(i\eta -z )^{2}} dz + 2/a.
\label{Reg2}
\end{equation}
This mathematical identity is employed to subtract from the integrand of $I_2$ a function
\begin{equation}
F_2(x) = h_0 \left[ {\rm Re}\frac{1}{[-g'_0(x-x_0)+i\eta ]^{2}}  + \frac{1}{[g'_0a]^{2}} \right],
\label{Reg3}
\end{equation}
with zero integral on the $(x_0-a,x_0+a)$ interval. The procedure leads to the following (suitable for Monte Carlo sampling) transformation in which one can simply put $\eta=0$ and omit the real part sign:
\begin{equation}
\rm{Re} \bar{R}_2 =\left\{
\begin{array}{ll}
         [R_2(x)  + R_2(x_r)-2F_2(x)]/2   & \mbox{if}\; |x-x_0|<a \\
          R_2(x)                          & \mbox{otherwise}.
\end{array} \right.
\label{R2b}
\end{equation}

Similarly, real part of the third-order pole is regularized using identity
\begin{equation}
\int_{-\infty}^{\infty} {\rm Re}\frac{z}{(i\eta -z)^{3}} dz = 0 ,
\label{Reg4}
\end{equation}
to suggest a function with zero integral on the $(x_0-a,x_0+a)$ interval
\begin{equation}
F_3(x) = H_0 \left[ {\rm Re}\frac{x-x_0}{[-g'_0(x-x_0)+i\eta ]^{3}}  - \frac{1}{a^2[g'_0]^{3}} \right],
\label{Reg3}
\end{equation}
where $H_0= h'_0-3h_0g''_0/2g'_0$. The value of $H_0$ is set by the requirement that $\rm{Re} \bar{R}_3$
\begin{equation}
\rm{Re} \bar{R}_3 =\left\{
\begin{array}{ll}
         [R_3(x)  + R_3(x_r)-2F_3(x)]/2   & \mbox{if}\; |x-x_0|<a \\
          R_3(x)                          & \mbox{otherwise}
\end{array} \right.
\label{R2b}
\end{equation}
is not singular at the pole and, thus, one can safely put $\eta=0$ in this expression. We do not elaborate on expressions for higher order poles after establishing the framework for their construction because diagrams considered in this work only feature poles with $n \le 2$.

%%%%%%%%%%%%%%%%%%%%%%%%%%%%%%%%%%%%%%%%%%%%%%%%%%%%%%%%%%%%%%%%%%%%%%%%%%%%%%%

\noindent \textit{Diagrammatic setup.} Diagrammatic series for HEG can be constructed in a number of ways all of which are supposed to produce identical results. In practice, setups producing accurate results already at relatively low expansion orders are selected and in this work we employ the one used in Refs.~\cite{BSE2024,LD2024}.
The expansion is performed in the number of screened interaction potentials $W_{st}(q)=V(q)/[1 - V(q)\Pi_{st}(q)]$, where $\Pi_{st}$ is the static RPA polarization of the ideal gas at zero temperature (the so-called Lindhard function \cite{Lindhard54}). Correspondingly, in high-order diagrams $\Pi_{st}\equiv V^{-1} - W^{-1}_{st}$ serves as a counter term to polarization loops based on the convolution of two Green's functions \cite{ShiftAct}. Similarly, bare Green's functions are replaced  with finite-$T$ solutions of the self-consistent HF approximation in all diagrams, $G^{-1}_{HF} = i\omega_m - \varepsilon_{HF}(k)$, where $\varepsilon_{HF}(k)=k^2/2m + \Sigma_F(k)-\mu$ is the HF dispersion relation, and $\Sigma_F(k) = \sum_{\mathbf q} W_{st}(\mathbf{q}) G_{HF}(\mathbf{k}-\mathbf{q}, \tau=-0)$ is the proper Fock self-energy (Hartree diagram is absent by charge neutrality), which also serves as a counter term for the lowest-order self-energy diagrams.

\begin{figure}[t]
\centerline{\includegraphics[width=0.95\columnwidth]{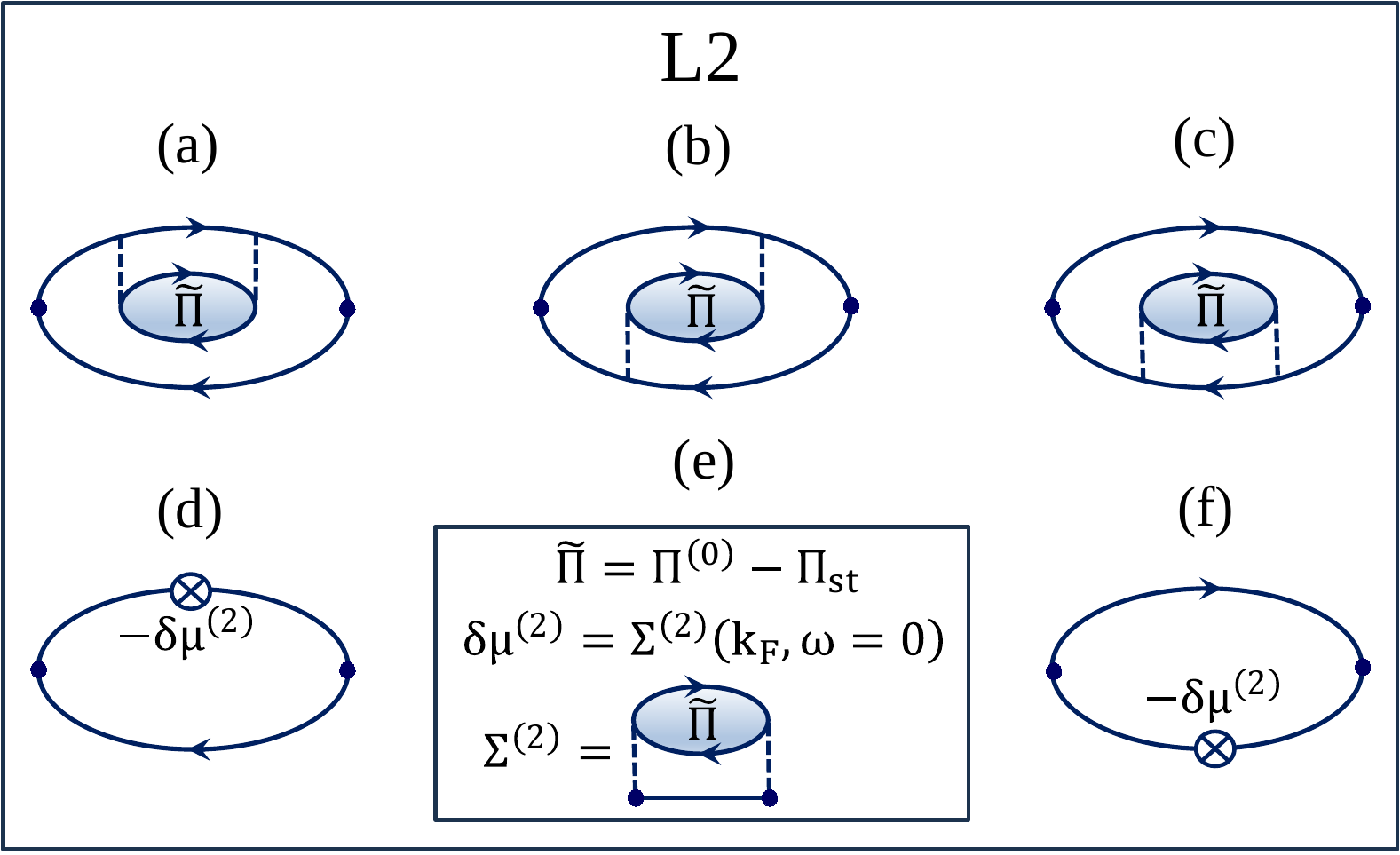}}
\caption{(a)-(c): Two-loop contributions to system's polarization with internal polarization loops containing counter-terms $\Pi_{st}$, see text and the definition of $\tilde{\Pi}$ in the inset (e). (d)-(f): Chemical potential counter-terms compensating for density changes caused by the proper self-energy insertion $\Sigma^{(2)}$ in diagrams (a) and (c), see inset (e). }
\label{Fig2}
\end{figure}

The HF-BSE approximation accounts for all polarization diagrams in the from of ladder series dressing the vertex. To go one step further, we add two-loop diagrams (taking care of required counter-terms) shown in Fig.~\ref{Fig2}. Diagrams $(a)-(c)$ contain contributions from two (e-h) pairs and, correspondingly, account for plasmon decay---they need to be considered together in order to satisfy the exact hydrodynamic condition $\Pi (Q\to 0, \Omega \ne 0) \propto Q^2$ and keep the plasmon frequency at zero momentum unchanged, $\omega_{pl}^2(Q=0) \equiv \Omega_{pl}^2 = 4 \pi e^2 \rho /m$. Finally, given that diagrams (a) and (b) are based on the Green's function renormalization, one has to consider chemical potential counter terms, see diagrams (d) and (f) in Fig.~\ref{Fig2}, in order to keep the Fermi momentum (and density) fixed. We abbreviate the new scheme as BSE+L2. [Up to change in notations, the analytic expressions for two-loop diagrams were considered in the Supplemental material for Ref.~\cite{TTKP2021}.]

%%%%%%%%%%%%%%%%%%%%%%%%%%%%%%%%%%%%%%%%%%%%%%%%%%%%%%%%%%%%%%%%%%%%%
\smallskip

\noindent \textit{Energy loss function.} The prime goal of this work is to demonstrate that regularization technique can be used for bias-free computations of real-frequency contributions to polarization from diagrams featuring poles with $n>1$, i.e. Monte Carlo simulations within the BSE+L2 scheme are numerically exact and do not introduce errors beyond the discussed diagrammatic approximations. The scheme is capturing physics of energy losses in the whole range of frequencies below $2 \div 3$ Fermi energies while keeping the computational cost low.
\begin{figure}[t]
\centerline{\includegraphics[width=0.95\columnwidth]{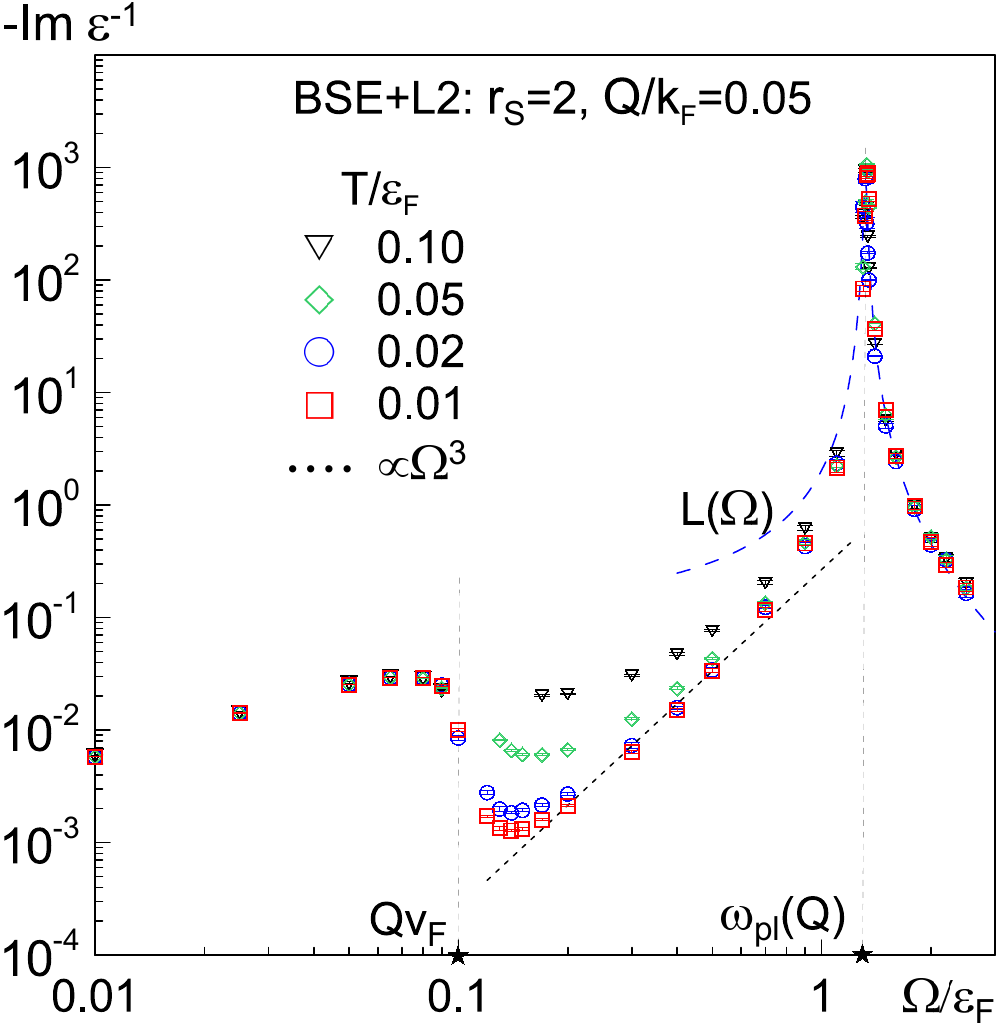}}
\caption{Energy loss function at $r_S=2$ and $Q/k_F=0.05$ as predicted by the BSE+L2 scheme. Curves with symbols correspond to different temperatures. The blue dashed line is a Lorentzian (unnormalized) with the half-width $\Gamma$ computed in Ref.~\cite{TTKP2021} at $T/\varepsilon_F=0.02$. The black dotted line is the $(\Omega/\varepsilon_F)^3$ law. Error bars are within the symbol sizes.}
\label{Fig3}
\end{figure}

In Fig.~\ref{Fig3} we present results for $-{\rm Im} \epsilon^{-1}(Q,\Omega,T)$ within the HEG model at $r_S=2$ and small momentum $Q/k_F = 0.05$. They have to be compared with predictions of the HF-BSE scheme shown in Fig.~\ref{Fig1} for the same parameters. As expected, the gap between the single $(e-h)$ pair continuum and undamped plasmon peak is now populated with two $(e-h)$ pairs. The thermal effects for two-pair excitations are surprisingly strong at frequencies $\Omega \gtrsim Q v_F$ to the extent that they eliminate nearly all traces of the single-pair threshold already at $T/\varepsilon_F \sim 0.1$. At much lower temperature, the two-pair contribution first emerges from the single-pair continuum threshold as the $\propto \Omega^3$ power law, which directly follows from the phase-volume considerations (there was a misprint in Ref.~\cite{RIXS2020} mentioning $\Omega^2$ dependence at low frequencies instead of the observed $\Omega^3$ one). It quickly increases in magnitude and half-way to the plasmon frequency exceeds the strength of the maximum single-pair signal. In other words, the notion of the spectral gap at frequencies $\Omega < \omega_{pl}$
becomes meaningless. Finally, on approach to the plasmon peak the ELF curve increases by several orders of magnitude and for $ |\Omega - \omega_{pl}| \ll \omega_{pl}$ is well described by the Lorentzian peak shown by the dashed line in Fig.~\ref{Fig3}. Its half-width saturates to a finite value at $T\to 0$, which agrees with the result computed in Ref.~\cite{TTKP2021} using small but finite $\eta$. [Calculations in Ref.~\cite{TTKP2021} took about $20$ times longer for the smallest $\eta$ considered than within the BSE+L2 scheme.]
\begin{figure}[t]
\centerline{\includegraphics[width=0.96\columnwidth]{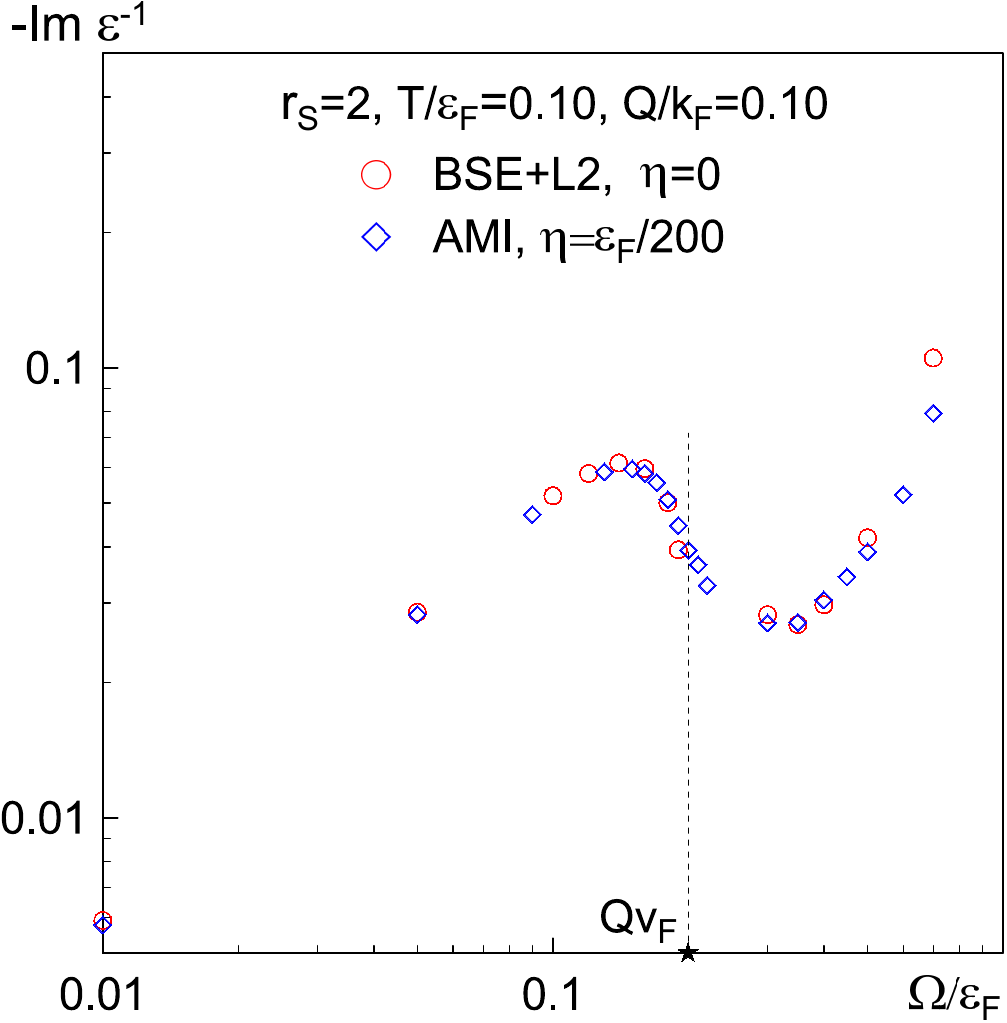}}
\caption{Comparison of energy loss functions for $r_S=2$, $T/\varepsilon_F=0.1$, and $Q/k_F=0.1$ obtained within the BSE+L2 technique (red circles) and the full diagrammatic expansion up to the third diagrammatic order inclusively using AMI with $\eta=T/20$ (blue diamonds)~\cite{LCHPT2022}.  }
\label{Fig4}
\end{figure}

At temperatures $T/\varepsilon_F > 0.02$ diagrams (a)-(c) produce visible spurious oscillations in the narrow frequency range around $Qv_F$. They are absent in Fig.~\ref{Fig3} because we omit data in this narrow region to avoid misinterpretation of results. Similar oscillations are observed in the order-by-order contributions within the HS-BSE scheme and smooth final HF-BSE results emerge only after series (re)summation. We thus expect oscillations to disappear after higher-order diagrams are accounted for. Our confidence is based on direct comparison between the ELF curves obtained within the regularized BSE+L2 scheme and the high-order AMI diagrammatic expansion with small $\eta = T/20 = \varepsilon_F/200$, see Fig.~\ref{Fig4}. [For the highest temperature in the set, it is possible to consider $\eta /T$ ratio small enough to obtain bias-free results but the corresponding AMI simulations are orders of magnitude more demanding in terms of the CPU time). Despite substantial difference in the diagrammatic setup and expansion orders considered, the two data sets for ELF are remarkably close to each other. This benchmark comparison establishes that the real-frequency BSE+L2 technique adequately captures the physical picture of energy losses in a wide range of frequencies while remaining computationally inexpensive (about a day on a desktop for the entire curve).

%%%%%%%%%%%%%%%%%%%%%%%%%%%%%%%%%%%%%%%%%%%%%%%%%%%%%%%%%%%%%%%%%%%%%

\noindent \textit{Conclusions.} Solutions of the ladder-type Bethe-Salpeter equation in the Hartree-Fock basis constitute an accurate and efficient approach for obtaining polarization at frequencies $\Omega \lesssim Qv_F$ where the dominant contribution comes from single electron-hole pairs. To extend it to higher frequencies while keeping the proper balance between the accuracy and efficiency, one has to consider two-loop diagrams capturing contributions from two electron-hole pairs and plasmon decay processes. The main technical obstacle in evaluating these additional diagrams on the real-frequency axis without systematic bias are the second-order pole singularities of the integrand. We explained how these (and higher-order) poles can be regularized in a manner compatible with efficient Monte Carlo sampling of momentum integrals. The resulting BSE+L2 scheme takes the limit $\eta \to 0$ after Wick's rotation $i\Omega \to \Omega + i\eta$ analytically.

New approach was applied to study the energy loss function of the homogeneous electron gas in the broad frequency range $\Omega \in [0, 5\varepsilon_F/2]$. The spectral gap between the single electron-hole excitations and plasmon peak existing at the level of RPA and HF-BSE approximations is transformed into the two electron-hole pair continuum featuring several important properties: (i) thermal effects at frequencies $\Omega \gtrsim Qv_F$ are strongly enhanced and almost completely erase signatures of the single-pair continuum already at $T/\varepsilon_F=0.1$;
(ii) at low temperature the energy loss function at $\Omega > v_F Q$ follows the $\propto \Omega^3$ law expected from the phase-volume considerations \cite{RIXS2020}; (iii) plasmon decay processes transform the $\delta$-functional peak into the Lorentzian line-shape, which accurately describes our data all the way up to the end of the considered frequency range. At the highest simulated temperature, the benchmark comparison between the BSE+L2 and high-order
diagrammatic expansion based on AMI with small but finite $\eta \ll T$ established that the new scheme is quantitatively accurate.

These developments suggest that exact pole regularization procedures allow one to compute contributions to polarization (or any other response function) on the real-frequency axis at finite temperature for an arbitrary Feynman diagram. The complexity of the calculation is increasing with the diagram order because for each pole the integrand is now represented by several, instead of one, functions but the evaluation of the integral is much more efficient because it does not suffer from divergent statistical measures and variance as $\eta \to 0$.

%%%%%%%%%%%%%%%%%%%%%%%%%%%%%%%%%%%%%%%%%%%%%%%%%%%%%%%%%%%%%%%

\smallskip

\noindent \textit{Acknowledgements.} We are grateful to David Huse for an inspiring discussion. This work was supported by the U.S. Department of Energy, Office of Science, Basic Energy Sciences, under Award DE-SC0023141.

\end{document}